\documentclass[12pt]{article}
\usepackage{amsmath}
\usepackage{multirow}
\usepackage{graphicx}
\usepackage{enumerate}
\usepackage{natbib}
\setcitestyle{aysep={}}
\usepackage{caption}
\usepackage{titlesec}
\usepackage{url} 
\usepackage{titlesec}

\newtheorem{prop}{Proposition} 

\newcommand{\blind}{0}

\begin{document}

\titleformat{\section}{\normalfont\Large\bfseries}{\thesection.}{1em}{}

\def\spacingset#1{\renewcommand{\baselinestretch}%
{#1}\small\normalsize} \spacingset{1}


\if1\blind
{
  \title{ }
  \author{Author 1\thanks{
    The authors gratefully acknowledge \textit{please remember to list all relevant funding sources in the unblinded version}}\hspace{.2cm}\\
    Department of YYY, University of XXX\\
    and \\
    Author 2 \\
    Department of ZZZ, University of WWW}
  \maketitle
} \fi

\if0\blind
{
  \bigskip
  \bigskip
  \bigskip
  \begin{center}
    {\LARGE\bf Composite Scores for Transplant Center Evaluation: A New Individualized Empirical Null Method}
\end{center}
  \medskip
} \fi

\begin{center}
    {Nicholas Hartman$^{1,2}$, Joseph M. Messana$^{2,3}$, Jian Kang$^{1,2}$, Abhijit S. Naik$^{2,3}$, Tempie H. Shearon$^{1,2}$, and Kevin He$^{1,2,*}$}
\end{center}
\smallskip
\begin{center}
    {$^1$Department of Biostatistics, University of Michigan, Ann Arbor, MI 48109} \\
    {$^2$Kidney Epidemiology and Cost Center, University of Michigan, Ann Arbor, MI 48109} \\
    {$^3$Division of Nephrology, University of Michigan, Ann Arbor, MI 48109}
\end{center}
\smallskip
\begin{center}
$^*email$: kevinhe@umich.edu
\end{center}

\bigskip
\begin{abstract}
\noindent 
Risk-adjusted quality measures are used to evaluate healthcare providers while controlling for factors beyond their control. Existing healthcare provider profiling approaches typically assume that the risk adjustment is perfect and the between-provider variation in quality measures is entirely due to the quality of care. However, in practice, even with very good models for risk adjustment, some between-provider variation will be due to incomplete risk adjustment, which should be recognized in assessing and monitoring providers. Otherwise, conventional methods disproportionately identify larger providers as outliers, even though their provider effects need not be ``extreme.'' Motivated by efforts to evaluate the quality of care provided by transplant centers, we develop a composite evaluation score based on a novel individualized empirical null method, which robustly accounts for overdispersion due to unobserved risk factors, models the marginal variance of standardized scores as a function of the effective center size, and only requires the use of publicly-available center-level statistics. The evaluations of United States kidney transplant centers based on the proposed composite score are substantially different from those based on conventional methods. Simulations show that the proposed empirical null approach more accurately classifies centers in terms of quality of care, compared to existing methods. 
\end{abstract}

\noindent%
\vfill

\newpage
\spacingset{1.0} 
\section{INTRODUCTION}
\label{sec:intro}

To improve quality of care and reduce costs for patients, the Centers for Medicare and Medicaid Services (CMS) monitors Medicare-certified healthcare providers nationwide with various quality measures of patient outcomes \citep{ash2012whitepaper,Liu2012,He2013, Kalbfleisch2013, Estes2018, Chen2019, estes2020profiling}. This monitoring can help patients make more informed decisions about where to receive care, and can also aid stakeholders in identifying providers that need to improve. In some cases, policymakers will intervene and fine or shut down providers with extremely poor outcomes. Thus, given the high stakes of provider profiling efforts, it is important that the statistical methods used in these applications produce accurate evaluations. 

Our  endeavor  is  motivated  by  the  study  of End-Stage Renal  Disease  (ESRD), which has substantial impacts on patient survival and quality of life. In 2019, there were 809,103 prevalent cases of ESRD in the United States (U.S.), which was an increase of 107.3\% since 2000 \citep{USRDS2021}. The only two treatments for ESRD are dialysis and transplantation, and  it has been shown that patients who receive kidney transplants have significantly better long-term prognoses than patients who remain on dialysis, so transplantation is the preferred treatment for eligible ESRD patients \citep{Wolfe}. However, despite aggressive efforts to increase the number of kidney donors, less than 13.7\% of incident ESRD patients in 2018 received a kidney transplant within one year, and the median wait time to receive a kidney transplant is four years \citep{USRDS2021}. At the same time, 20\% of donated kidneys are discarded \citep{Marrero}. To optimize treatment strategies for ESRD patients, it is important to encourage providers to perform more transplants and reduce organ waste.

Considering the important roles that transplant centers play in patients' post-transplant outcomes, the CMS and other stakeholders have made substantial efforts to monitor the quality of care provided by these centers. Historically, kidney transplant centers have been evaluated based on one-year patient and graft survival rates, and while post-transplant outcomes are crucial aspects of healthcare quality, there is growing concern that this evaluation method has incentivized centers to only perform transplants on patients with the least amount of risk \citep{Jay}. That is, to avoid higher mortality rates, centers may perform very few transplants and discard viable organs \citep{Hart2020}, limiting access to transplantation.  Based on this reasoning, there is a need for evaluation methods that balance post-transplant outcomes with other important components of healthcare, such as access to transplantation. Under a request by the CMS to develop such evaluation methods which can simultaneously assess multiple aspects of care, we propose a composite score that combines post-transplant outcome measures with a diverse set of other healthcare quality measures, summarizing distinct aspects of care into an overall assessment score. 

\section{SRTR DATASET AND STATISTICAL ISSUES}
\label{sec:gaps}

\subsection{SRTR Dataset}

The data for this application come from the Scientific Registry of Transplant Recipients (SRTR), which routinely publishes transplant center performance statistics in the form of Program Specific Reports (PSRs). The PSRs from the August, 2020 publication cycle for 212 kidney transplant centers are used in our analyses. We select four individual quality measures to form the proposed composite score: Transplant Rate Ratio (TRR), Standardized Acceptance Ratio (SAR), One-Year Patient Standardized Mortality Ratio (PSMR), and One-Year Standardized Graft Failure Ratio (GSMR). The first two measures reflect access to transplantation and the last two reflect patient outcomes after transplantation. Detailed descriptions of the measures are available in online Appendix A and the SRTR's technical documentation \citep{SRTR_psr}. 

\subsection{Statistical Issues}
\label{sec:gaps2}

The quality measures that contribute to a composite score are usually based on risk-adjustment models, which attempt to control for factors that are unrelated to a center's quality of care \citep{Shwartz,Proudlove}. Existing approaches typically assume that the risk adjustment is perfect and the between-provider variation is entirely due to the quality of care, which are often invalid assumptions \citep{Jones2011, Kalbfleisch2018, He2019PIUR}. In practice, even with very good models for risk adjustment, there will be risk factors that are not completely accounted for (e.g. unobserved socio-economic factors and comorbidities), and many of these characteristics will be related to the outcomes and vary across providers. Thus, some of the between-provider variation in a quality measure will be due to this incomplete risk adjustment.

For example, transplant centers depend on Organ Procurement Organizations (OPOs) to coordinate organ donations effectively \citep{OConnor}. If a transplant center resides in a service area that is operated by a poorly-performing OPO, it could appear to have a low transplant rate through partial fault of the OPO. However, in practice, it may be difficult to adjust for all aspects of the relationship between an OPO and a transplant center. In fact, the SRTR does not include any variables related to OPO performance in its risk-adjustment models \citep{SRTR_risk}. 

\begin{figure}[h!]
\vspace{-0.5cm}
    \centering
\includegraphics[width=\textwidth]{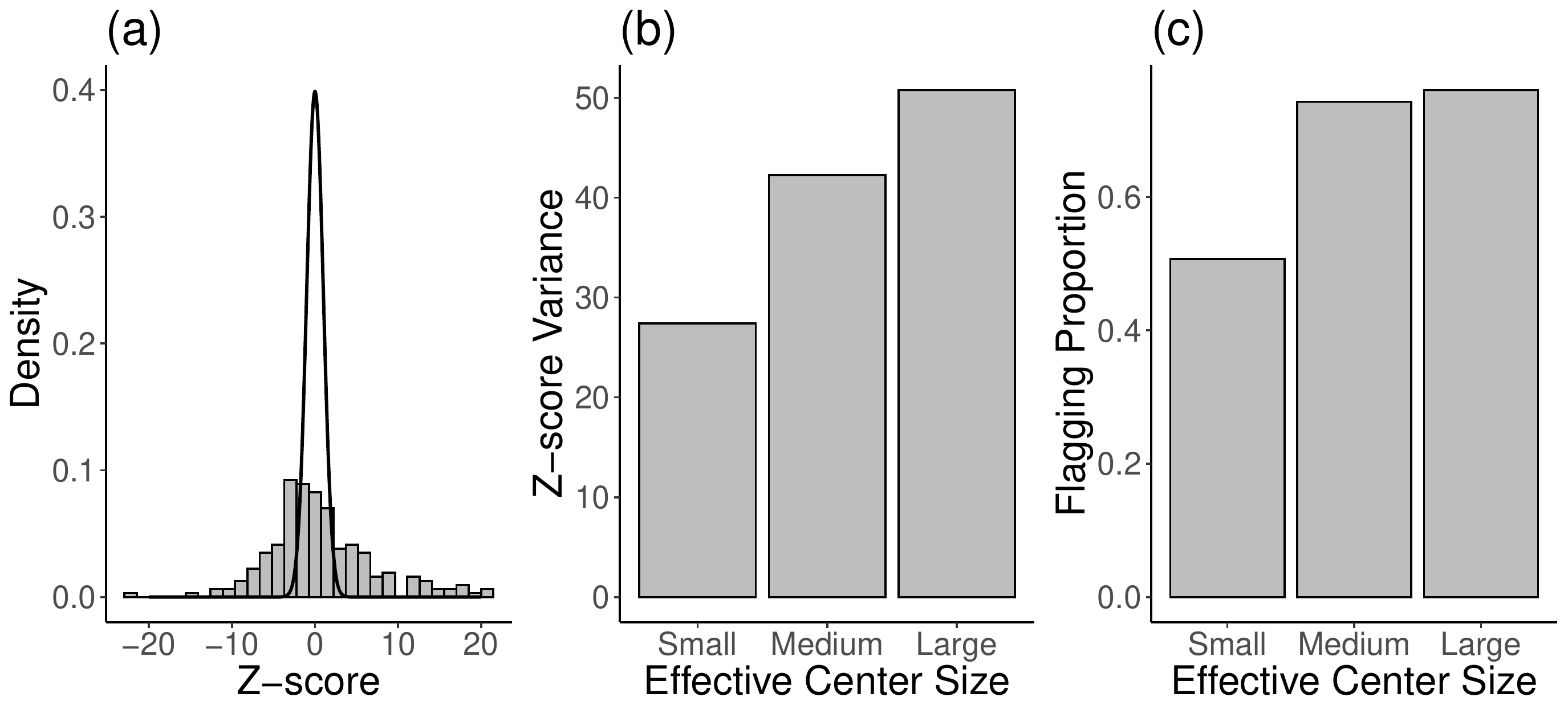} \caption{(a): Histogram of Transplant Rate Ratio Z-scores; Solid curve: standard normal distribution. (b): Variance of Z-score within each group of transplant centers. (c): Proportion of centers flagged for providing extremely poor or good care.}
 \label{fig:transplant}
\end{figure}

\begin{figure}[h!]
\vspace{-0.5cm}
    \centering
\includegraphics[width=\textwidth]{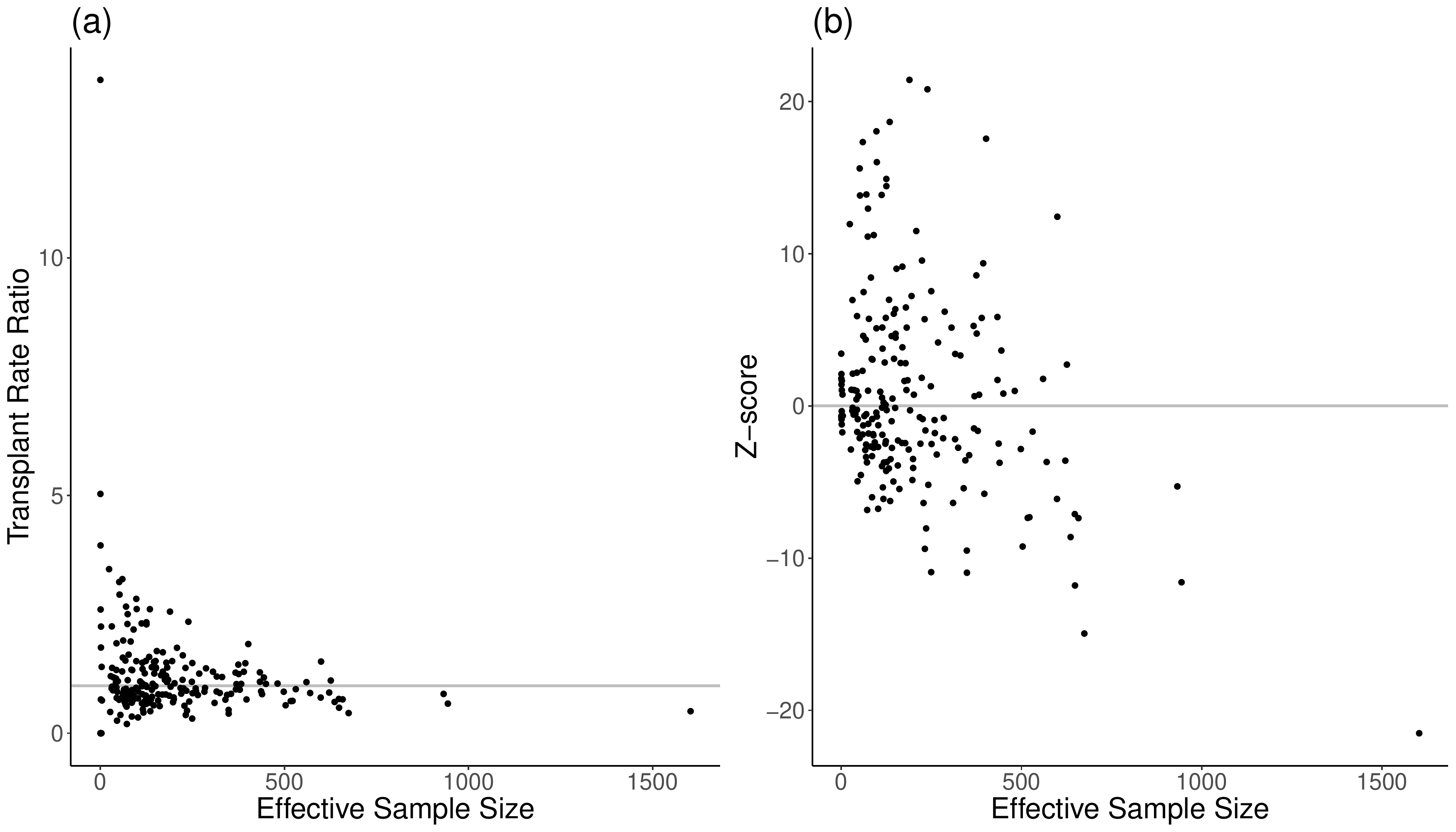} \caption{Plots of (a) Transplant Rate Ratio and (b) standardized Z-score against effective sample size. The solid line corresponds to the null value.}
 \label{fig:transplant2}
\end{figure}
 
To illustrate this problem, Figure \ref{fig:transplant} shows Z-scores corresponding to the TRR measures of the 212 centers from the SRTR dataset. If risk adjustment were perfect, the empirical distribution of the Z-scores would closely resemble the standard normal distribution. However, as shown in Figure \ref{fig:transplant}, the empirical distribution is much more dispersed than the standard normal distribution. If we perform a hypothesis test (at level 0.05) for each center by comparing the Z-scores to a standard normal distribution, two-thirds of the centers are flagged for providing care that is significantly different from the national average. When we divide the centers  into three  groups  based  on  their effective sample sizes (i.e. the expected numbers of transplants given the centers' patient characteristics), the  empirical variance of the  Z-scores and the proportion of flagged centers increases across groups as the average effective sample size increases. Figure \ref{fig:transplant2} shows that centers with large effective sample sizes can have Z-scores that are very far from the null value, even if the measure values are close to the null value and not clinically significant. That is, conventional methods disproportionately flag larger centers for providing extremely poor or good care, when many of these centers in reality provide average care and should not be penalized or rewarded. 

Several existing methods have been proposed for modeling overdispersion due to unobserved confounders in provider profiling. One common approach is to estimate a multiplicative scale factor, which can be used to correct the variance of the Z-scores under the null hypothesis \citep{Spiegelhalter_Plot}. These methods typically apply the same corrective factor to each provider's Z-score, and, hence, disproportionately flag large providers (e.g. Figure \ref{fig:transplant}). A few existing methods that allow the overdispersion correction to depend on the effective sample size suffer from their own limitations. For example, the method-of-moments proposed by \citet{Spiegelhalter_Plot} estimates corrective factors for overdispersion using all available centers in the dataset, including those with very extreme Z-score values. Thus, due to overestimation of the overdispersion factor, its flagging power tends to be compromised in the presence of outliers. We demonstrate this property through numerical evaluations in Section \ref{sec:compare}. In addition, \citet{Xia2020} proposed a method that involves grouping the centers by sample size and smoothing the overdispersion corrections within the groups. However, the choice of grouping is arbitrary and using a different number of groups will change the list of flagged centers. Moreover, in our application, there are only 212 centers, so it is impossible to form enough groups and have enough data within the groups to implement this method.

Another challenge in working with the SRTR database is that, due to patient privacy, direct sharing of sensitive individual information is restricted for some quality measures. For example, the risk-adjustment models for the SAR are based on detailed donor-level information corresponding to offered kidney donations. However, the CMS claims and administrative databases only contain donor-level information for kidneys that become accepted by the center for transplantation. Therefore, methods that require the availability of patient-level data for improving model fitting, such as those under the causal inference framework \citep{Angrist}, are not suitable for our application.

To fill these gaps, we propose the following methods: First, in recognition of the fact that unobserved confounding can cause overdispersion in conventional scoring systems, we propose an individualized empirical null approach to transform individual quality measures so that they follow similar scales and can be combined into a single score. The proposed method accounts for the unexplained variation between centers and is robust to outlying centers. Moreover, the method robustly models the between-center variance as a function of effective sample size and avoids the bias against large centers. Second, we combine individual quality measures into a composite score, using an approach that down-weighs highly correlated metrics. This approach prevents redundant information from dominating the composite score, which would hinder our efforts to balance distinct aspects of care in our assessments. The proposed method only requires center-level summary statistics, and no sensitive patient-level data are needed. Thus, it is applicable to broader settings when direct sharing of sensitive individual data is restricted. 

\section{METHODS}
\label{sec:method}

\subsection{Composite Score Framework}
\label{sec:overview}

Let $F$ be the number of transplant centers under evaluation and let $P$ be the number of distinct quality measures recorded for each center. Denote $i$ as the center index where $i=1,\dots,F$ and $k$ as the measure index where $k=1,\dots,P$. The patient or donation records within the $i^{th}$ center are indexed by $j=1,\dots,n_{ik}$, where $n_{ik}$ is the number of records for the $k^{th}$ measure, within the $i^{th}$ center.

Construction of a composite score from $P$ distinct measures involves two key steps: standardization and combination. As discussed above, the fundamental motivation for the standardization step is incomplete risk adjustment. That is, the estimated measures are subject to overdispersion from unobserved confounders, which disproportionately affects larger centers (as exemplified in Figure 1). Moreover, depending on the type of available data, researchers may be able to perform more complete risk adjustment for some outcomes compared to others. Thus, the amount of overdispersion within each measure must be taken into account to ensure fair evaluations of the centers. 

Another key motivation for standardization is that quality measures may be derived from different types of outcomes (such as count, binary, and survival outcomes) and may be estimated with different levels of precision. Therefore, the individual measures are calculated on very different scales. Standardization maps the measures onto comparable scales and facilitates combination of the individual measures.

\subsection{Model}
\label{sec:GLM}

Assume that the observed outcome for the $i^{th}$ center, $j^{th}$ patient, and $k^{th}$ measure, $Y_{ijk}$, is generated from a Generalized Linear Model (GLM) and has the following density from the exponential family of distributions, \begin{equation}
\label{eq:glm}
\pi_k(Y_{ijk}; \theta^*_{ijk}, \psi_k) \propto \exp\bigg (\frac{ Y_{ijk}\theta^*_{ijk}-b_k(\theta^*_{ijk})}{a_k(\psi_k)}\bigg), \end{equation}

\noindent where $a_k(\cdot)$ and $b_k(\cdot)$ are pre-specified functions that correspond to the distributional assumption for $Y_{ijk}$, and $\psi_k$ is a nuisance parameter. It follows from GLM theory that $E[Y_{ijk}|\theta^*_{ijk}, \psi_k]=b'_k(\theta^*_{ijk})$ and $Var[Y_{ijk}|\theta^*_{ijk}, \psi_k]=a_k(\psi_k)b''_k(\theta^*_{ijk})$. The canonical parameter, $\theta^*_{ijk}$, is $$
\theta^*_{ijk}=\mu_k+\gamma^*_{ik}+\alpha_{ik}+\boldsymbol{x_{ijk}^T\beta_k},$$


\noindent where $\mu_k$ is the population norm, $\gamma^*_{ik}$ is the effect of the center's quality of care, $\boldsymbol{\beta_k}$ is a vector of coefficients, $\boldsymbol{x_{i1k}}, \ldots, \boldsymbol{x_{in_{ik}k}}$  are independently and identically distributed realizations of a random  vector $\boldsymbol{X}_{ik}$ that corresponds to patient-level covariates, and $\alpha_{ik}$ is a random quantity corresponding to omitted variables, with $E[\alpha_{ik}|\boldsymbol{X}_{ik}]=0$ and  $Var[\alpha_{ik}|\boldsymbol{X}_{ik}]=\sigma_{\alpha_k}^2$.

We assume that $\mu_k$ and $\boldsymbol{\beta_k}$ are estimated with enough precision such that they can be treated as fixed quantities. This assumption is common in provider profiling applications, which usually involve datasets with very large sample sizes \citep{Kalbfleisch2013, Jones2011}. For applications with smaller datasets, the proposed methods can be adapted to account for the variability in these estimates, using Bayesian or empirical Bayes methods \citep{Jones2011,Carlin}.

For $i=1,\dots,F$ and the $k^{th}$ individual measure, our goal is to test the null hypotheses $H_{0ik}^*: \gamma^*_{ik}=0$. A statistic is constructed for each center to determine if the quality of care is significantly different from the national average (conditional on a set of patient and center characteristics), with respect to the $k^{th}$ type of outcome.

\subsection{Naive Fixed-Effects Standardization}
\label{sec:standard}

Since $\alpha_{ik}$ is unobserved, a naive testing approach is to replace $\theta^*_{ijk}$ with a working version, $\theta_{ijk}$, which is defined as $\theta_{ijk}=\mu_k+\gamma_{ik}+\boldsymbol{x_{ijk}^T\beta_k}$, where $\gamma_{ik}=\gamma^*_{ik}+\alpha_{ik}$. Under this misspecified model, the effect of the unobserved center-level confounder is absorbed into the effect of quality of care. The only testable null hypothesis under this misspecified model is $H_{0ik}: \gamma_{ik}=0$, and a naive standardized Z-score can be constructed for each center using a fixed-effects score test: \begin{equation}
\label{eq:FE}
    Z_{FE,ik}=\frac{\sum_{j=1}^{n_{ik}}\{Y_{ijk}-b'_k(\theta^0_{ijk})\}}{\sqrt{a_k(\psi_k)\sum_{j=1}^{n_{ik}} b''_k(\theta^{0}_{ijk})}},
\end{equation}

\noindent where $\theta^0_{ijk}=\mu_k+\boldsymbol{x_{ijk}^T\beta_k}$. 

While the above testing approach is based on a misspecified model, it is still frequently used by stakeholders in provider profiling applications. In the context of transplant center evaluations, detailed patient-level data on kidney donors from the general population are restricted, due to data privacy standards. Thus, we can only use center-level statistics from the SRTR that are estimated from working models which are likely misspecified. Since we do not have access to sufficient patient-level data, we cannot improve the estimates from the misspecified models directly, so we must correct the naive Z-scores using center-level information only. 

\subsection{Empirical Null Framework}
\label{sec:approach}

If $\theta^0_{ijk}$ were correctly specified (e.g. the risk adjustment is perfect), $Z_{FE,ik}$ would asymptotically follow a standard normal distribution under the null hypothesis. However, since perfect risk adjustment is unrealistic and $\alpha_{ik}$ is unobserved, the working model is misspecified, and hence, $Z_{FE,ik}$ no longer follows an asymptotic standard normal distribution when $\gamma^*_{ik}=0$. In online Appendix B, we derive an approximate function for the null variance, which we summarize below in Proposition 1.

\begin{prop}
   \label{prop}
  Under Model (\ref{eq:glm}) and the null hypotheses $H_{0ik}^*: \gamma^*_{ik}=0$, given $\sum_{j=1}^{n_{ik}}b'_k(\theta^0_{ijk})$ and $\widetilde{n}_{ik}$,
  $Z_{FE,ik}/\sqrt{1+{\varphi}_k\widetilde{n}_{ik}}$
 asymptotically follows a standard normal distribution, where $\varphi_k=\sigma^2_{\alpha_k}/a_k(\psi_k)$ and $\widetilde{n}_{ik}=\sum_{j=1}^{n_{ik}} b''_k(\theta_{ijk}^0)$, which is the effective center size.
\end{prop}

Thus, the null variance of $Z_{FE,ik}$ is larger than one and approximately a linear function of the effective center size.  Note that conventional methods for simultaneous inference typically assume that the data are generated from identical distributions and thus use a pooled empirical null estimate \citep{Efron, efron2007size, efron2010large}. 
However, unlike classical settings in which sample sizes are the same for all tests, here center volumes and the influences of unmeasured confounders vary across centers, resulting in center-specific null distributions.

Based on the above result, we develop an empirical null method that estimates the variance of $Z_{FE,ik}$ for each center under the null hypothesis. We model the density of the observed $Z_{FE,ik}$ values as a mixture of null and non-null densities: $$
    \pi_{0k}f_{0,ik}(z|\widetilde{n}_{ik})+(1-\pi_{0k})f_{1,ik}(z),$$

\noindent where $\pi_{0k}$ is the proportion of centers that provide average care with respect to the $k^{th}$ measure. In practice, it is assumed that $f_{0,ik}(z|\widetilde{n}_{ik})$ corresponds to the normal density with mean zero and variance $1+\varphi_k\widetilde{n}_{ik}$. The non-null density, $f_{1,ik}(z)$, is left unspecified, but we require that its support is outside of some pre-specified interval, which  ensures that the parameters of the null distribution are identifiable. This assumption is useful in the context of provider profiling because we are mainly interested in identifying outlying centers that have clinically meaningful differences from the population norm. Under this framework, the parameter $\varphi_k$ can be estimated through maximum likelihood estimation (MLE) methods, and we describe the details of this estimation procedure in Section \ref{sec:estimation}.

After obtaining $\widehat{\varphi}_k$, the MLE of $\varphi_k$, we construct Z-scores based on the empirical null by performing a simple correction of the original Z-scores: \begin{equation}
\label{eq:Z_null}
    Z_{EN,ik}=\frac{Z_{FE,ik}}{\sqrt{1+\widehat{\varphi}_k\widetilde{n}_{ik}}}.
\end{equation}

\noindent If the outcome data are generated from Model (\ref{eq:glm}), then $Z_{EN,ik}$ asymptotically follows a standard normal distribution under the null hypothesis, even if $\alpha_{ik}$ is unobserved. Table~\ref{tab:stats} provides a summary of the center-level statistics that are needed to implement our empirical null method for common outcome variable distributions. Researchers can perform evaluations by obtaining the public center-level statistics described in Table~\ref{tab:stats}, without having to use patient-level information. As described in Section \ref{sec:gaps}, this aspect of the proposed method is practically useful, since organ acceptance measures are based on sensitive and restricted data from organ donors in the general U.S. population. In contrast, the CMS claims and administrative databases, which are our main data sources for profiling, only contain sufficient data on accepted kidney donations and do not represent the general donor population. Because our method only requires center-level summary statistics, we can overcome these data restrictions. 

\begin{table}[h]
\centering
\caption{Center-level summary statistics required to compute Z-scores from the individualized empirical null method with outcome distributions in the exponential family. For the Normal distribution with an identity link function, $\sigma^2_{\varepsilon_k}$ is the variance of the error term. For the Binomial distribution with a logit link function, $p_{ijk}^0={\exp(\mu_k+\boldsymbol{x_{ijk}^T\beta_k})}/{\{1+\exp(\mu_k+\boldsymbol{x_{ijk}^T\beta_k})\}}$. The individualized empirical null correction is a direct function of columns 3-5, which are all at the center level.}
\label{tab:stats}
\begin{tabular}{ccccc}
\hline
\textbf{Distribution} & \textbf{Link Function} & $\boldsymbol{\widetilde{n}_{ik}}$ &  $\boldsymbol{a_k(\psi_k)}$ & 
$\boldsymbol{\varphi_k}$  \\
 \hline
 Normal & Identity & $n_{ik}$ & $\sigma^2_{\varepsilon_k}$ & $\sigma^2_{\alpha_k}/\sigma^2_{\varepsilon_k}$ \\
 Binomial & Logit & $\sum_j p^0_{ijk}(1-p^0_{ijk})$ & 1 & $\sigma^2_{\alpha_k}$ \\
 Poisson & Log & $\sum_j \exp(\mu_k+\boldsymbol{x_{ijk}^T\beta_k})$ & 1 & $\sigma^2_{\alpha_k}$ \\
 \hline
\end{tabular}
\end{table}

\begin{figure}[h!]
\vspace{-0.5cm}
    \centering
\includegraphics[width=\textwidth]{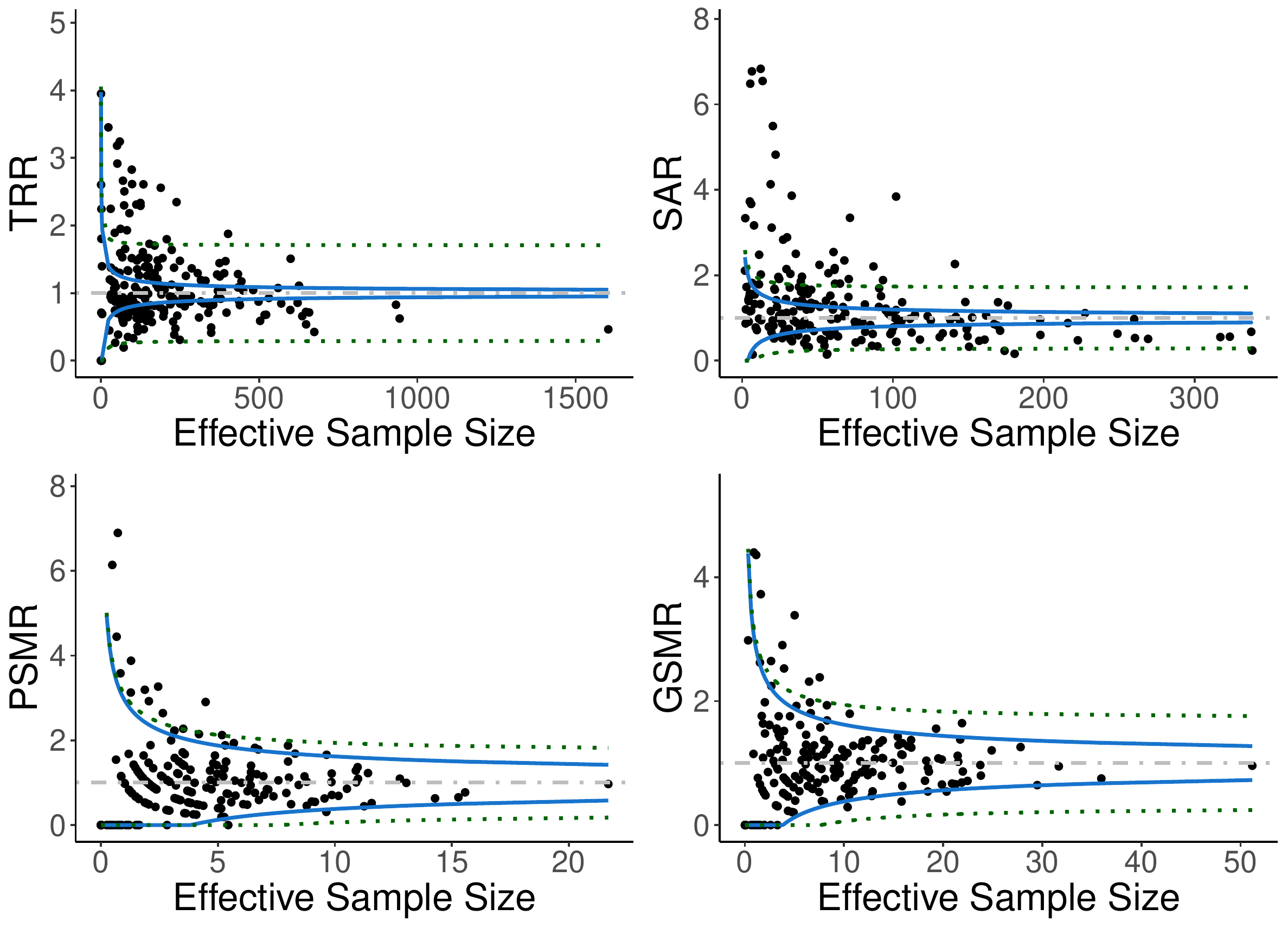} \caption{Funnel plots for four measures, using different standardization methods. TRR: Transplant Rate Ratio, SAR: Standardized Acceptance Ratio, PSMR: Patient Standardized Mortality Ratio, GMSR: Standardized Graft Failure Ratio. Centers that fall outside of the control limit lines are considered outliers. The solid and dotted lines correspond to the control limits based on fixed-effects standardization and empirical null standardization, respectively. The dot-dashed line represents the null value of one. One center with extremely large measure values was excluded for visual clarity.}
 \label{fig:plot}
\end{figure}

The funnel plots in Figure \ref{fig:plot} show the control thresholds for the four measures used in our composite score, which is described in detail in Section \ref{sec:real}. The differences in the control thresholds demonstrate the impact of the empirical null standardization. The thresholds based on fixed-effects standardization are very close to the null value, which means that a large number of centers fall outside of the thresholds and are thus labeled as outliers. From the theoretical arguments above, we suspect that many of these centers are flagged due to unobserved confounding effects which are unrelated to quality of care. The empirical null method flags fewer centers than fixed-effects standardization, and the centers that are flagged by this method have more clinically meaningful measure values. 

\subsection{Empirical Null Estimation}
\label{sec:estimation}

Estimation of $\varphi_k$ is achieved through MLE methods under a truncated normal model. We assume that the support of $f_{1,ik}(z)$ is defined outside of the interval $[A_{ik},B_{ik}]$ where $A_{ik}=-v \sqrt{1+\widehat{\varphi}^0_k\widetilde{n}_{ik}}$, $B_{ik}=v \sqrt{1+\widehat{\varphi}^0_k\widetilde{n}_{ik}}$, $v$ is a selected constant (such as the 95$^{th}$ or 97.5$^{th}$ percentile of the standard normal distribution), and $\widehat{\varphi}^0_k$ is an initial estimate of $\varphi_k$. The initial estimate, $\widehat{\varphi}^0_k$, can be generated from robust M-estimation methods \citep{Huber}. Let $I_{0k}=\{i:\ Z_{FE,ik} \in [A_{ik}, B_{ik}]\}$ and define $\phi_{\varphi_k}(z|\widetilde{n}_{ik})$ as the normal density with mean zero and variance $1+\varphi_k\widetilde{n}_{ik}$. If we denote $Q_{ik}(\varphi_k)$ as the probability that $Z_{FE,ik}$ falls in the interval $[A_{ik},B_{ik}]$ under the null hypothesis, then the likelihood function can be written as
\begin{equation}
\label{eq:Like}
 L(\varphi_k, \pi_{0k}) =  \prod_{i \in I_{0k}} \{\displaystyle {\pi_{0k}\phi_{\varphi_k} (Z_{FE,ik}|\widetilde{n}_{ik})}\}
 \prod_{i \notin I_{0k}}
 \{1 - \pi_{0k}Q_{ik}(\varphi_k)\}.
\end{equation}

\noindent For a fixed $\pi_{0k}$, we maximize~(\ref{eq:Like}) with respect to $\varphi_k$ through the Nelder-Mead algorithm, with $\widehat{\varphi}^0_k$ serving as the initial solution \citep{NM}. An MLE of the null proportion is then obtained through a grid search algorithm on the profile likelihood function. Further details are provided in online Appendix C.

\subsection{Weighting and Flagging}
\label{sec:flagging}

After standardization, the measures are combined to form the final composite score; the most popular approach is to take a weighted sum of the individual standardized measures. The choice of weight function is non-trivial, and many options have been explored \citep{Shwartz}. From the resulting composite scores, researchers can flag a subset of the centers as extremely poor or extremely good overall performers, which is the main goal of composite score development. 

Without loss of generality, we assume that a lower value of $Z_{EN,ik}$ corresponds to worse quality of care, for $k=1,\dots,P$. If this assumption does not hold, researchers can easily multiply certain individual scores by negative-one to achieve this interpretation. After computing $Z_{EN,ik}$ for $k=1,\dots,P$, the composite score is constructed by taking a weighted sum, using the following form \citep{Spiegelhalter}: \begin{equation}
\label{eq:sum}
Z_{CS,i}=\left(\sum_{k=1}^P\sum_{\ell=1}^Pw_kw_{\ell}c_{k\ell}\right)^{-1/2}\sum_{k=1}^P w_k Z_{EN,ik}, \end{equation}

\noindent where $w_k$ is the weight for the $k^{th}$ standardized score, $c_{kl}$ is the estimated correlation between the $k^{th}$ and $\ell^{th}$ standardized scores, and the $CS$ in $Z_{CS,i}$ stands for the composite score. Since $Z_{EN,ik}$ asymptotically follows a standard normal distribution under the null hypothesis, the first term in the right side of (\ref{eq:sum}) ensures that $Z_{CS,i}$ does as well. 

There are many possible options for the weight function in (\ref{eq:sum}). For example, researchers may select the weights to reflect a priori knowledge of each measure's clinical importance. On the other hand, \citet{Spiegelhalter} proposed a weight function that assigns less weight to highly-correlated measures, which prevents redundant information from contributing too much to the composite score. The function is $w_k=\sum_{\ell=1}^P t_{k\ell}$, where $t_{kl}$ is the element in row $k$ and column $\ell$ of the inverse correlation matrix, $\boldsymbol{T}=\boldsymbol{C^{-1}}$. This weighting scheme is an attractive option because it has theoretical justification. However, we have found in our analyses that this weight function can be either positive or negative, which may result in problematic clinical interpretations of the composite score. 

\citet{Spiegelhalter} also noted that for applications with many quality measures, the computation of $\boldsymbol{T}$ may be infeasible. To overcome these computational difficulties, they proposed an alternative weight function that no longer requires matrix inversion and is defined as $w_k=1/\sum_{\ell=1}^P \textrm{max}(c_{k\ell},0)$. In our motivating application of transplant center evaluation, the computation of $\boldsymbol{T}$ is feasible, so we would be able to use the original weight function proposed by \citet{Spiegelhalter}. However, this alternative method guarantees that the weights are positive, so we use it in our application. 

Once the composite score is computed, some centers are flagged for providing care that deviates from the national average. Since $Z_{CS,i}$ asymptotically follows a standard normal distribution under the null hypothesis, a natural flagging approach is to perform a hypothesis test on each center by comparing the observed $Z_{CS,i}$ to the standard normal distribution. Specifically, a control limit or threshold can be selected that corresponds to a certain significance level. 

In practice, we may use thresholds of -1.96 and 1.96, which correspond to the 5\% significance level. Any center with $Z_{CS,i} < -1.96$ or $Z_{CS,i} > 1.96$ is flagged as an extremely poor or good provider, respectively. In the context of transplant center evaluation, it is more important to detect and flag problematic centers than it is to strictly control the family-wise error rate, as long as a reasonably low number of average centers are flagged. Therefore, the use of multiple testing corrections is rarely justified in this setting, as these methods can reduce the power of our flagging approach. 

\section{U.S. TRANSPLANT CENTER EVALUATIONS}
\label{sec:real}

\subsection{Analysis Approach}

We first computed $Z_{FE,ik}$ values from (\ref{eq:FE}) for each of the 212 centers and the four measures, and we estimated the empirical null distributions from these values. The TRR, PSMR, and GSMR are based on count data, so we assumed that the corresponding outcomes for these measures follow Poisson distributions. The SAR is based on binary observations, so we assumed a Bernoulli distribution for the outcome variable. For all measures, we computed $Z_{EN,ik}$ values from (\ref{eq:Z_null}). There were low correlations between the access to transplantation measures and the patient outcome measures, but high correlations among the two measures in each of these classes. The quality measures that reflect access to transplantation had larger $\widehat{\sigma}^2_{\alpha_k}$ estimates compared to the measures that reflect patient outcomes, suggesting that there was greater overdispersion for these measures (Table~\ref{tab:sig}). We constructed the composite scores using the $Z_{EN,ik}$ values and the correlation-based weights in (\ref{eq:sum}), with $w_k=1/\sum_{\ell=1}^P \textrm{max}(c_{k\ell},0)$. 

\begin{table}[h!]
    \centering
    \caption{Sample correlation matrix, correlation-based weights, and estimates of the unobserved confounding effect variance for measures used in the composite score. } 
    \label{tab:sig}
    \begin{tabular}{ccccccc}
    \hline
     & \multicolumn{4}{c}{\textbf{Sample Correlations}} & & \\
     \cline{2-5}
      & TRR & SAR & PSMR & GSMR &  \textbf{Weight} & $\boldsymbol{\widehat{\sigma}^2_{\alpha_k}}$ \\
     \hline
      TRR  & 1.00 & 0.64 & 0.03 & -0.02 &  0.39 & 0.14 \\
      SAR & 0.64 & 1.00 & -0.02 & -0.03 & 0.40 & 0.24 \\
      PSMR & 0.03 & -0.02 & 1.00 & 0.73 & 0.37 & 0.04 \\
      GSMR & -0.02 & -0.03 & 0.73 & 1.00 & 0.38 & 0.04 \\
      \hline
    \end{tabular}
\end{table}

\subsection{Flagging Results and Comparisons}

After applying threshold-based flagging (with control limits of -1.96 and 1.96), we flagged 4.2\% of centers for providing extremely poor overall care and 9.0\% of centers for providing extremely good care. We also compared these flagging results to those from a composite score based on $Z_{FE,ik}$ instead of $Z_{EN,ik}$. This version of the composite score does not correct for incomplete risk adjustment, and it flagged more than 50\% of centers for providing poor or good care (Table~\ref{tab:label}). 

Finally, we compared these results to those from method-of-moments standardization. As described in Section \ref{sec:gaps}, the main difference between the method-of-moments and the proposed empirical null is that the method-of-moments uses all centers to estimate the corrective factor, including potential outliers, which produces overly-conservative evaluations. The composite score based on method-of-moments standardization only flagged 1.4\% of centers for providing poor care (Table \ref{tab:label}). In contrast, the individualized empirical null uses the two-group model in (\ref{eq:Like}) and is more robust to outliers.

\begin{table}[h!]
    \centering
    \caption{Transplant center flagging results based on composite scores with different methods of standardization. The percentages of centers that are labeled as providing poor, average, or good care. EN: Empirical null.} 
    \label{tab:label}
    \begin{tabular}{cccc}
    \hline
     \textbf{Standardization Type} &  \textbf{Poor Care (\%)} & \textbf{Average Care (\%)} & \textbf{Good Care (\%)} \\ 
     \hline
     Fixed-Effects & 25.94  & 45.75 & 28.30 \\
     Method-of-Moments & 1.42 & 91.98  & 6.60 \\
     Individualized EN & 4.24 & 86.79 & 8.96 \\
      \hline
    \end{tabular}
\end{table}

\begin{figure}[h!]
\vspace{-0.5cm}
    \centering
\includegraphics[width=\textwidth]{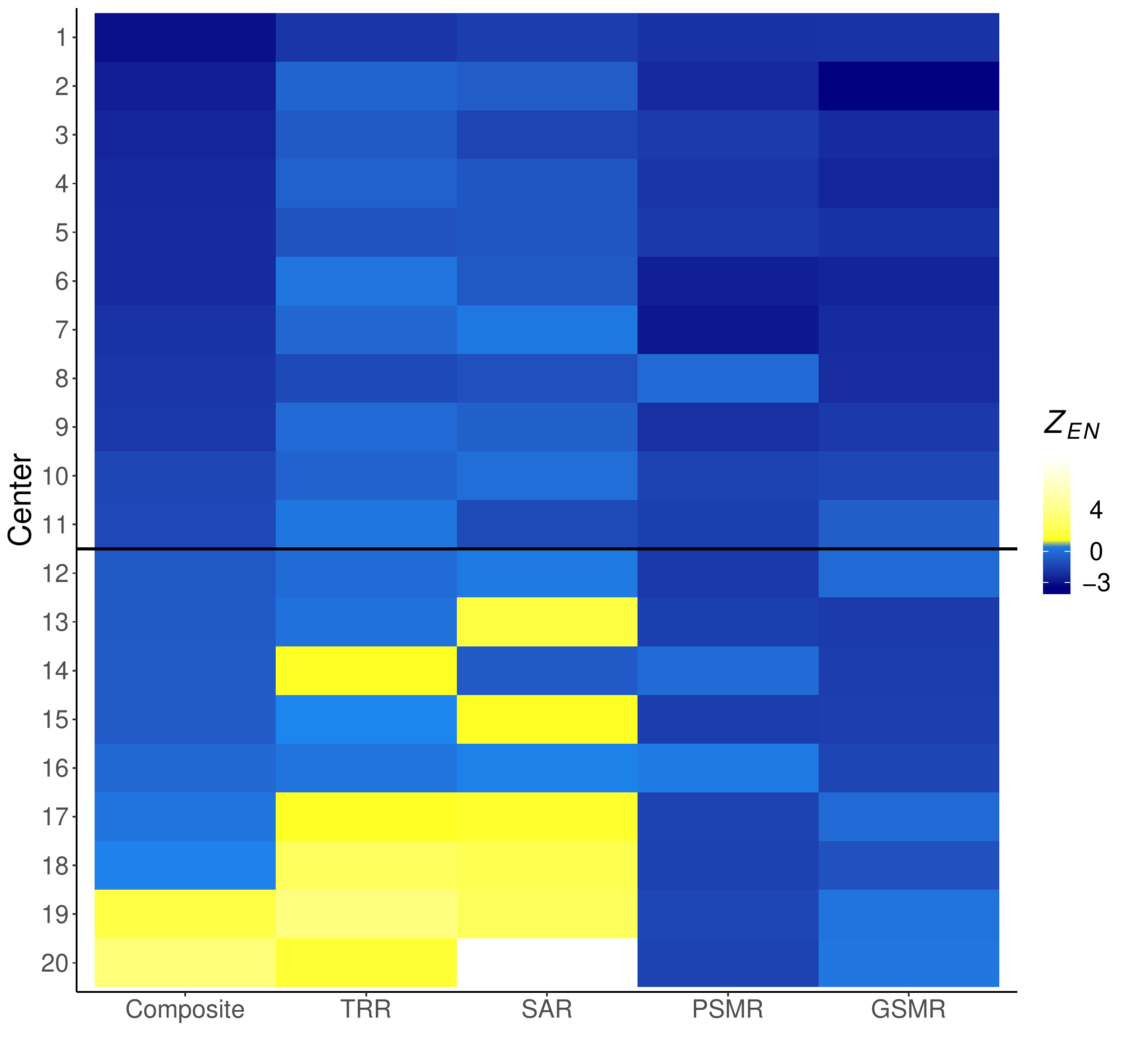} \caption{Heat map of center performance, for the 20 centers that would be flagged as poor-performers using historical methods (based on the PSMR and GSMR measures only). Lower $Z_{EN}$ values (and darker shades) correspond to worse quality of care for the corresponding measures. Centers that fall above the horizontal line are flagged as poor-performers by the proposed composite score approach. TRR: Transplant Rate Ratio, SAR: Standardized Acceptance Ratio, PSMR: Patient Standardized Mortality Ratio, GMSR: Standardized Graft Failure Ratio.}
 \label{fig:heat}
\end{figure}

\begin{figure}[h!]
\vspace{-0.5cm}
    \centering
\includegraphics[width=\textwidth]{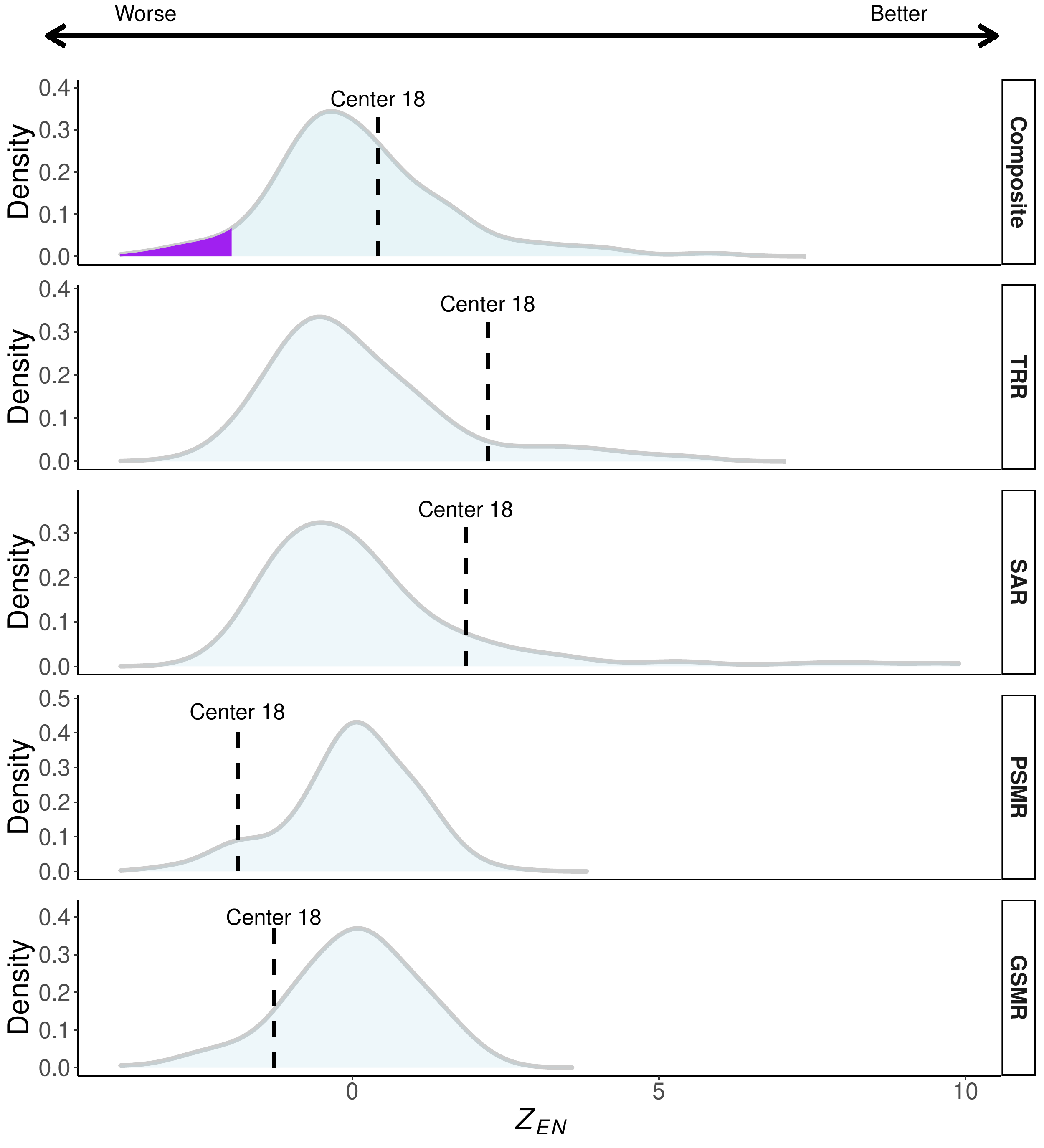} \caption{National distributions of the standardized Z-scores ($Z_{EN}$) for the composite score and each measure component. Lower Z-score values correspond to worse quality of care for the corresponding measures. The Z-score for Center 18 from Figure \ref{fig:heat} is marked with a dashed line. The dark shaded region of the top panel represents the flagging region based on a threshold of -1.96. TRR: Transplant Rate Ratio, SAR: Standardized Acceptance Ratio, PSMR: Patient Standardized Mortality Ratio, GMSR: Standardized Graft Failure Ratio.}
 \label{fig:center18}
\end{figure}

Figure \ref{fig:heat} shows a visual heat map of the quality of care among those centers that would be flagged as poor-performers, using historical flagging methods. Since the historical flagging rule is based only on post-transplant outcomes, the centers that are flagged under this rule have poor performance with respect to the PSMR or GSMR. In Figure \ref{fig:heat}, some of these centers are not flagged under the composite score approach, and this subset of centers tends to have better performance in terms of access to transplantation. For example, Center 18 has high PSMR and GSMR values, but it performs exceptionally well in terms of accepting organs at a high rate and transplanting many patients. In other words, this center is successful in expanding access to transplantation, which inevitably involves the transplantation of some high-risk patients and thus can result in higher mortality rates. In Figure \ref{fig:center18}, which shows the performances of Center 18 relative to the national distributions, it is evident that the composite score balances patient outcomes with access to transplantation.

The objective of the proposed composite score method is to encourage better access to transplantation and to avoid the unintended negative incentives caused by only using post-transplant outcomes as the evaluation criteria. Therefore, it is undesirable to flag centers such as Center 18 strictly for having higher mortality rates. The composite score achieves this objective, since we observe that some centers with high transplant and acceptance rates can have favorable composite scores, even if the mortality rates are elevated compared to the national average. It should be noted that the composite score does not promote reckless transplantation, because post-transplant outcomes are still incorporated in the overall score. Rather, the composite score balances both of these aspects in the evaluation, whereas the historical rule is only based on post-transplant outcomes.

\section{SIMULATIONS}
\label{sec:compare}

\subsection{Individual Measure Simulation Design}
\label{sec:indiv}

We used simulations to compare the performances of three standardization methods (fixed-effects, method-of-moments, and the individualized empirical null). To simplify the comparisons, we first focused our assessments on a single  quality measure. The performances of the standardization methods for each individual measure have direct impacts on the final composite scores, which we demonstrate in Section \ref{sec:composite_sim}. The data were generated from Model (\ref{eq:glm}) for 212 transplant centers, with effective sample sizes that resembled the real SRTR dataset. We generated $x_{ijk}$ from a $\textrm{N}(-0.4,0.5)$ distribution and the number of person-years in the center-specific cohorts, $r_{ik}$, from an Exponential distribution with a mean of 1000. The model coefficient parameters were defined as $\mu_k=-6$ and $\beta=1$. These simulation settings were selected because they generated realistic datasets that were similar to the SRTR data. 

For one of the centers, which we denoted as ``Center 1," the quality of care effect, $\gamma^*_{1k}$, was varied from zero to four. In our first simulation setting, all remaining centers had $\gamma^*_{ik}=0$, and we referred to these as ``null centers." In a separate simulation setting, $\gamma^*_{ik}$ was set as equal to one for 5\% of the centers, and negative-one for an additional 5\% of the centers, which introduced centers with outlying quality of care effects.  On each iteration of the simulation, $\alpha_{ik}$ was generated from a $\textrm{N}(0,0.14)$ distribution. The $Y_{ijk}$ values were sampled from Poisson distributions with means $\exp(\theta^*_{ijk}+log(r_{ik}))$. All hypothesis testing was performed at the 5\% level, and simulation results were based on 1000 iterations. 

The individualized empirical null method depends on a selected constant $v$ to define the support of the non-null distribution (Section \ref{sec:approach}). In practice, we can let $v$ be the $(100-q)^{th}$ percentile of the standard normal distribution, where $q$ is a constant, such as five or ten. In addition, \cite{Spiegelhalter} suggests that the method-of-moments can be implemented using Winsorized Z-scores, meaning that any Z-score which is more extreme than the $q^{th}$ or $(100-q)^{th}$ percentile of the Z-score distribution is replaced with that percentile value. We assessed the robustness of the empirical null method and the method-of-moments to their respective tuning parameters by repeating the simulations above, with values of $q$ ranging from 2.5 to 15. 

\subsection{Individual Measure Simulation Results}

\begin{figure}[h!]
\vspace{-0.5cm}
    \centering
\includegraphics[width=\textwidth]{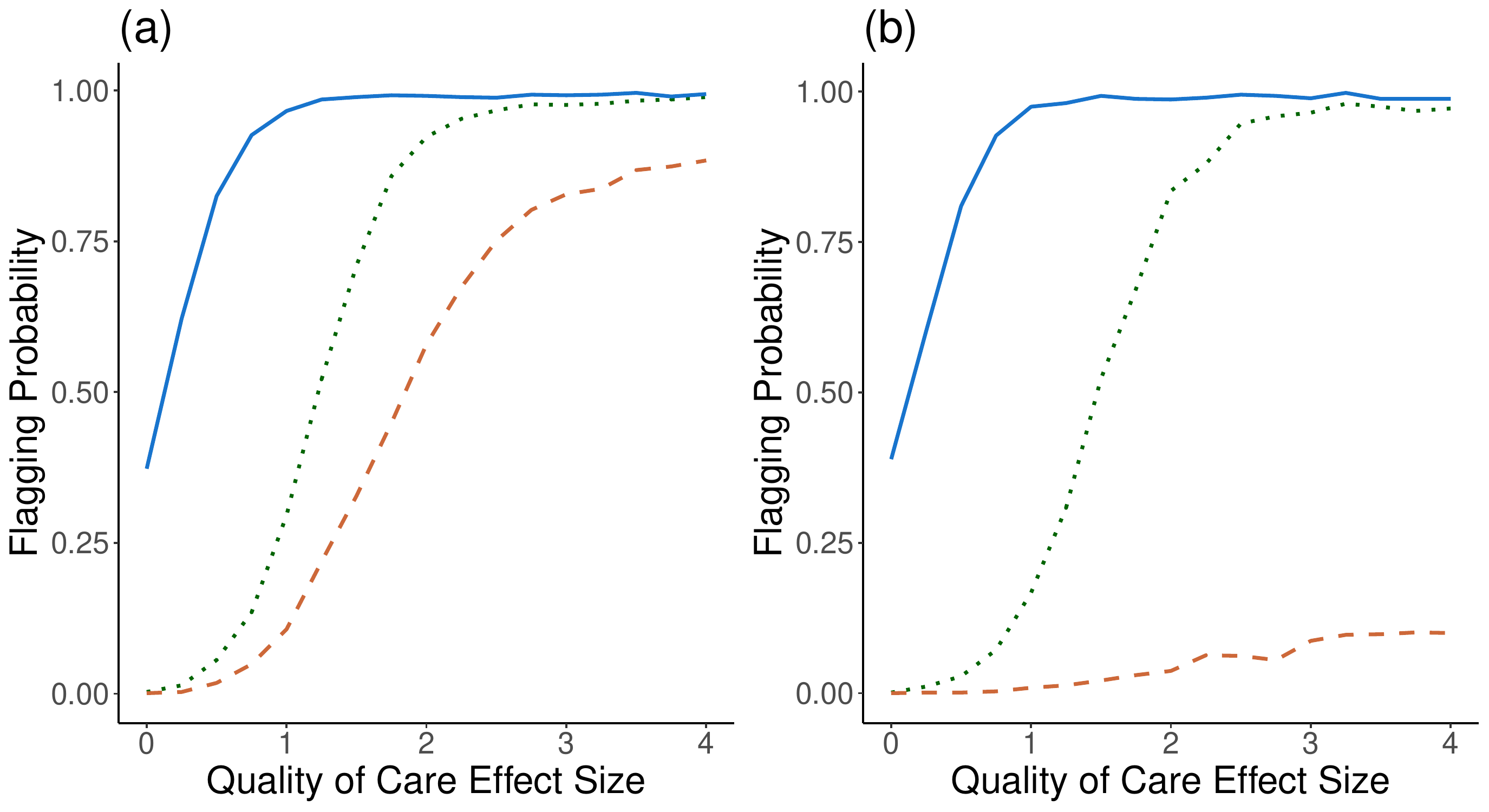} \caption{Flagging probability for different quality of care effect sizes ($\gamma^*_{ik}$), with (a) no outliers and (b) 10\% outliers. The solid, dashed, and dotted lines correspond to fixed-effects, method-of-moments, and empirical null standardization respectively. The fixed-effects approach does not account for overdispersion, the method-of-moments is an existing correction (Section \ref{sec:gaps2}), and the empirical null is the proposed correction.}
 \label{fig:out}
\end{figure}

\begin{figure}[h!]
\vspace{-0.5cm}
    \centering
\includegraphics[width=\textwidth]{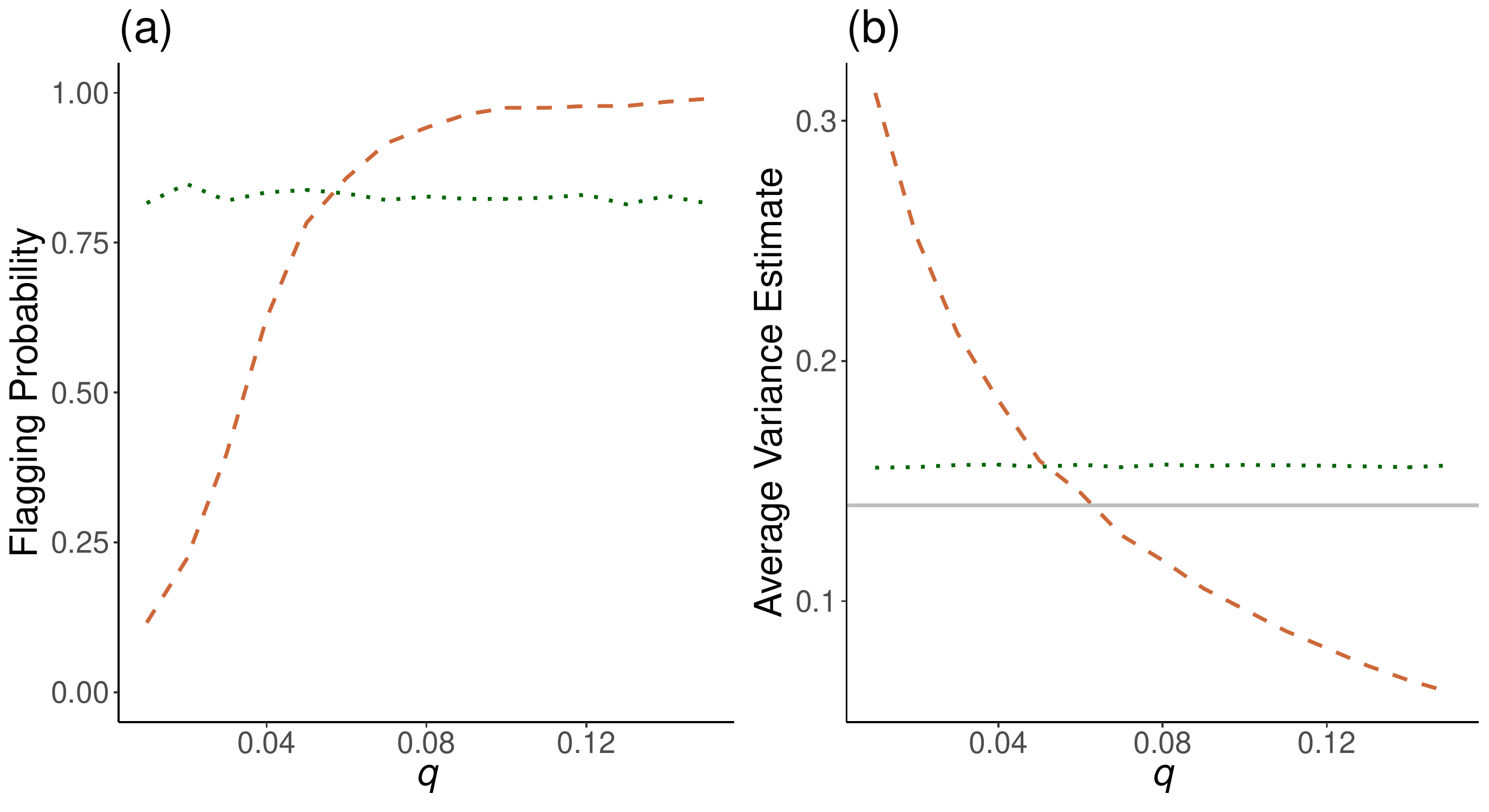} \caption{(a): Flagging probability and (b): Average estimate of $\sigma^2_{\alpha_k}$ (the variance of the unobserved quantity, $\alpha_{ik}$), for different $q$ values. The dashed and dotted lines correspond to method-of-moments and empirical null standardization respectively. In (b), the solid line corresponds to the true value of 0.14. The method-of-moments is an existing correction (Section \ref{sec:gaps2}), and the empirical null is the proposed correction.}
 \label{fig:win}
\end{figure}

The simulation results in Figures \ref{fig:out} and \ref{fig:win} highlight the limitations of existing standardization methods. When the quality of care effect size for Center 1 was set as equal to zero, fixed-effects standardization incorrectly flagged Center 1 with a high probability. On the other hand, the method-of-moments and the empirical null method had flagging probabilities that were close to zero under this setting, so they were both able to control the probability of flagging a null center. For all non-zero values of the quality of care effect size, the empirical null method had more power to flag the outlying center compared to the method-of-moments. In addition, the empirical null was robust to the introduction of other outlying centers, but the method-of-moments had a substantial decrease in power when there were other outlying centers (Figure \ref{fig:out}).

According to our simulations, another limitation of the method-of-moments is that it is more sensitive to the level of Winsorization than the empirical null method is to the choice of $v$ value (Figure \ref{fig:win}). When the method-of-moments was implemented with no Winsorization or a low level of Winsorization (i.e. a $q$ value near zero), it severely overestimated the variance of the unobserved quantity, $\alpha_{ik}$. As a result, the power of the method-of-moments was much lower for these settings compared to the empirical null. As the amount of Winsorization increased, there were dramatic changes in the bias of the unobserved quantity's variance estimate and in the amount of power to detect Center 1's outlying quality of care effect. The empirical null method had a stable flagging probability, and minimal bias in the variance estimation across all settings of $q$. 

\subsection{Composite Score Simulations}
\label{sec:composite_sim}

To study the impacts of different standardization methods on the composite score, we simulated multiple measures to be combined into a single score. For simplicity, we generated two independent measures from Poisson models, using the same simulation structure as in Section \ref{sec:indiv}. Higher measure values were interpreted as corresponding to better quality of care for the first measure and worse quality of care for the second measure. Thus, based on (\ref{eq:sum}), the composite score was computed as the difference in two Z-scores, divided by $\sqrt{2}$. In the context of our motivating application, the first measure would reflect access to transplantation and the second measure would represent adverse patient outcomes. This interpretation of the simulated measures is used throughout this section. We set $\sigma^2_{\alpha_1}=0.14$ and $\sigma^2_{\alpha_2}=0.04$, which are similar to the estimates we observed in our motivating dataset (Table \ref{tab:sig}). Data were generated for 212 centers, where 10\% of the centers were considered outliers for at least one of the measures and 90\% of the centers had $\gamma^*_{i1}=\gamma^*_{i2}=0$. 

We focused our analyses on the composite scores and flagging rates of four specific centers: Center 1 with poor access and poor outcomes ($\gamma^*_{11} < 0$ and $\gamma^*_{12}=-\gamma^*_{11}$), Center 2 with good access and good outcomes ($\gamma^*_{21} > 0$ and $\gamma^*_{22}=-\gamma^*_{21}$), Center 3 with poor access and good outcomes, and Center 4 with good access and poor outcomes. For Centers 3 and 4, $\gamma^*_{i1}$ was first set as equal to a non-zero constant, and $\gamma^*_{i2}$ was defined such that the expected composite score would be equal to zero, assuming all parameters are known (online Appendix D). For all four centers, the quality of care effect for the first measure ($\gamma^*_{i1}$) was varied from 0.1 to 1.0. Ideally, Centers 1 and 2 would be flagged with a high probability and Centers 3 and 4 would be flagged much less often.

\begin{figure}[h!]
\vspace{-0.5cm}
    \centering
\includegraphics[width=\textwidth]{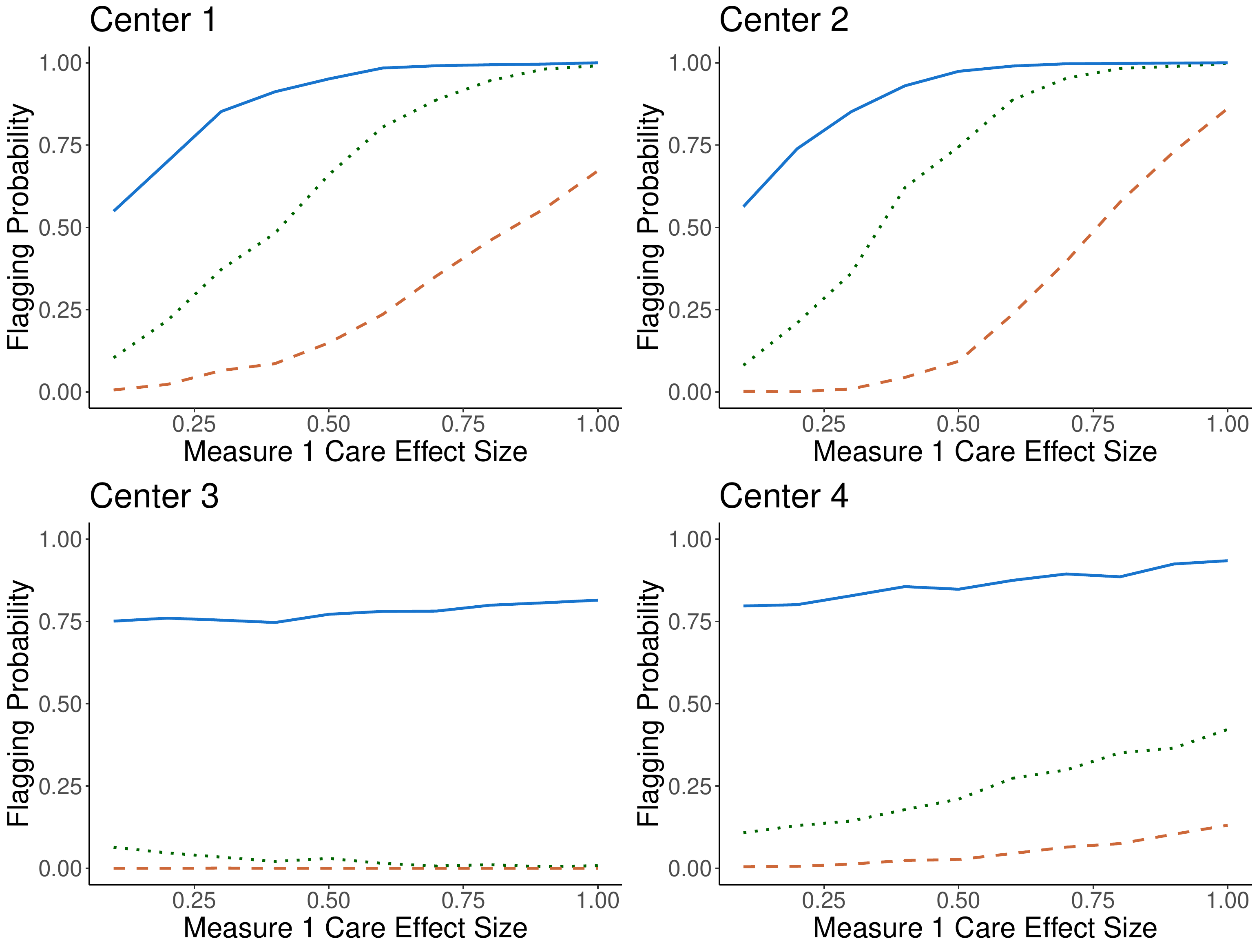} \caption{Flagging probabilities based on a simulated composite score of two measures for four centers of interest: Center 1 with poor access and poor outcomes, Center 2 with good access and good outcomes, Center 3 with poor access and good outcomes, and Center 4 with good access and poor outcomes. The horizontal axis corresponds to the quality of care effect size with respect to the first measure (i.e. $\gamma^*_{i1}$ in Section \ref{sec:GLM}). The solid, dashed, and dotted lines correspond to fixed-effects, method-of-moments, and empirical null standardization respectively. The fixed-effects approach does not account for overdispersion, the method-of-moments is an existing correction (Section \ref{sec:gaps2}), and the empirical null is the proposed correction.}
 \label{fig:combine}
\end{figure}

The composite score based on fixed-effects standardization flagged all four centers of interest with very high probability (Figure \ref{fig:combine}). Therefore, this method was not useful for identifying outlying centers in terms of overall quality of care. The composite score based on method-of-moments standardization had lower flagging rates for Centers 3 and 4, as desired, but it also had much less power to detect Centers 1 and 2 as outliers. The proposed composite score, based on empirical null standardization, also limited the flagging rates of Centers 3 and 4 well, and it had more power to detect Centers 1 and 2 as outliers, compared to the method-of-moments (Figure \ref{fig:combine}).

\section{DISCUSSION}
\label{sec:disc}

Motivated by the clinical importance of transplantation in ESRD patients, we have developed a composite evaluation score that incentivizes centers to provide better access to transplantation. The proposed score balances post-transplant outcomes such as mortality with access to transplant measures, which allows stakeholders to easily assess many distinct aspects of a center's quality of care. We argue that adoption of composite scores for transplant center evaluation could encourage centers to perform more transplants and hold centers accountable for organ waste.

To overcome the challenges in evaluating transplant centers and the limitations of existing standardization methods, we have proposed an individualized empirical null approach which models the variance of the null distribution as a function of the effective sample size. The empirical null method accounts for overdispersion due to unobserved confounders and can be implemented without using patient-level data. Compared to existing methods which either ignore overdispersion or use all centers to estimate a corrective factor, the individualized empirical null performs better in terms of controlling false positives and maintaining enough power to detect centers with extremely poor or good care. 

We note that the quality of care effects in Model (\ref{eq:glm}) can also be estimated under a random effects framework, which is an alternative to the fixed-effects test described in Section \ref{sec:method}. However, naive applications of random effects can lead to biased estimates of $\boldsymbol{\beta_k}$ when $\gamma^*_{ik}$ and $\boldsymbol{X_{ik}}$ are correlated or when there are outlying centers, whereas the fixed-effects estimates remain unbiased under these situations \citep{Kalbfleisch2013}. In addition, naive applications of random effect testing approaches fail to account for overdispersion due to unobserved confounding \citep{Jones2011}. Thus, the fixed-effects approach is preferred under our setting. In future work, we extend the individualized empirical null to a Bayesian framework, which accounts for potential variability in the MLE. 

\section*{ACKNOWLEDGEMENTS}

The authors would like to thank Dr. John D. Kalbfleisch for his helpful comments. The data that support the findings of this paper are publicly available from the Scientific Registry of Transplant Recipients at \url{https://www.srtr.org/reports/program-specific-reports/}. 

\section*{DISCLOSURE STATEMENT}

The authors report there are no competing interests to declare. 

\section*{FUNDING}

This work was partially supported by the Centers for Medicare and Medicaid Services under Contract 75FCMC21C0035 and the National Institute of Diabetes and Digestive and Kidney Diseases under Grant R01DK129539. The statements contained in this article are solely those of the authors and do not necessarily reflect the views or policies of the Centers for Medicare and Medicaid Services or the National Institute of Diabetes and Digestive and Kidney Diseases.

\renewcommand{\refname}{REFERENCES}

\bibliographystyle{JASA}
\bibliography{main}

@article{Spiegelhalter,
    author={Spiegelhalter, D. and Sherlaw-Johnson, C. and Bardsley, M. and Blunt, I.
    and Wood, C. and Grigg, O.},
    year={2012},
    title={{Statistical Methods for Healthcare Regulation: Rating, Screening and Surveillance}},
    journal={Journal of the Royal Statistical Society},
    volume={175},
    number={1},
    pages={1-47}
}

@article{He2013,
  title={{Evaluating Hospital Readmission Rates in Dialysis
Facilities; Adjusting for Hospital Effects}},
  author={He, K. and Kalbfleisch, J.D. and Li, Y. and Li, Y.J.},
  journal={Lifetime Data Analysis},
volume={19},
  number={4},
  pages={490--512},
  year={2013}
}

@article{Kalbfleisch2013,
  title={{On Monitoring Outcomes of Medical Providers}},
  author={Kalbfleisch, J.D. and Wolfe, R.A.},
  journal={Statistics in Biosciences},
    volume={5},
  number={2},
  pages={286--302},
  year={2013}
}

@article{Estes2018,
  title={{Time-Dynamic Profiling with Application to Hospital Readmission Among Patients on Dialysis}},
  author={Estes, J.P. and Nguyen, D.V. and Chen, Y. and Dalrymple, L.S. and Rhee, C.M. and Kalantar-Zadeh, K. and Sent\"{u}rk, D.},
  journal={Biometrics},
      volume={74},
  number={4},
  pages={1383--1394},
  year={2018}
}

@article{estes2020profiling,
  title={{Profiling Dialysis Facilities for Adverse Recurrent Events}},
  author={Estes, J.P. and Chen, Y. and {\c{S}}ent{\"u}rk, D. and Rhee, C.M. and K{\"u}r{\"u}m, E. and You, A.S. and Streja, E. and Kalantar-Zadeh, K. and Nguyen, D.V},
  journal={Statistics in Medicine},
  volume={39},
  number={9},
  pages={1374--1389},
  year={2020},
  publisher={Wiley Online Library}
}

@article{He2019PIUR,
author = {He, K. and Dahlerus, C. and Xia, L. and Li, Y. and Kalbfleisch, J.},
year = {2019},
title = {{The Profile Inter‐Unit Reliability}},
volume = {76},
journal = {Biometrics},
number={2},
pages={654–663}
}

@book{USRDS2021,
author={{United States Renal Data System}},
 title={{2021 USRDS Annual
Data Report: Epidemiology of Kidney Disease in the United States}},
  year={2021},
  publisher={National Institutes of Health, National Institute of Diabetes and Digestive and Kidney Diseases, Bethesda, MD}
}

@article{Xia2020,
    author = {Xia, L. and He, K. and Li, Y. and Kalbfleisch, J.},
    title = {{Accounting for Total Variation and Robustness in Profiling Health Care Providers}},
    journal = {Biostatistics},
    year = {2022},
    volume={23},
    number={1},
    pages={257–273}
}

@misc{ash2012whitepaper,
	title={{Statistical Issues in Assessing Hospital Performance}},
	author={Ash, A.S. and Fienberg, S.F. and Louis, T.A. and Normand, S.T. and Stukel, T.A. and Utts, J.},
	note={{\it Commissioned by the Committee of Presidents of Statistical Societies for the Centers for Medicare and Medicaid Services (CMS)} [online], Available at \url{https://www.cms.gov/Medicare/Quality-Initiatives-Patient-Assessment-Instruments/HospitalQualityInits/Downloads/Statistical-Issues-in-Assessing-Hospital-Performance.pdf}},
	year={2012}
}

@article{Jones2011,
  title={{The Identification of Unusual Health-Care Providers from a Hierarchical Model}},
  author={Jones, H.E. and Spiegelhalter, D.J.},
  journal={The American Statistician},
    volume={65},
  number={3},
  pages={154--163},
  year={2011}
}

@article{Kalbfleisch2018,
  title={{Does the Inter-Unit Reliability (IUR) Measure Reliability?}},
  author={Kalbfleisch, J.D. and He, K. and Xia, L. and Li, Y.M.},
  journal={Health Services and Outcomes Research Methodology},
     volume={18},
  number={3},
  year={2018},
  pages={215–225}
}

@article{Hart2020,
  title={{OPTN/SRTR 2018 Annual Data
Report: Kidney}},
  author={Hart, A. and Smith, J.M. and Skeans, M.A. and Gustafson, S.K. and Wilk, A.R. and Castro, S. and Foutz, J. and Wainright, J.L. and Snyder, J.J. and Kasiske, B.L. and Israni, A.K.},
  journal={American Journal of
Transplant},
    volume={20},
  number={1},
  year={2020}
}

@article{Chen2019,
  title={{Association of US Dialysis Facility Staffing with Profiling of Hospital-Wide 30-Day Unplanned Readmission}},
  author={Chen, Y and Rhee, C and Senturk, D and Kurum, E and Campos, L and Li, Y and Kalantar-Zadeh, K and Nguyen, D},
  journal={Kidney Diseases},
    volume={5},
  number={3},
  pages={153-162},
  year={2019}
}

@article{Liu2012,
  title={{Computationally Efficient Marginal Models for Clustered Recurrent Event Data}},
  author={Liu, D. and Schaubel, D.E. and Kalbfleisch, J.D.},
  journal={Biometrics},
    volume={68},
  number={2},
  pages={637-647},
  year={2012}
}

@article{OConnor,
title={{OPO Performance Improvement and Increasing Organ Transplantation: Metrics are Necessary but Not Sufficient}},
author={Kevin O’Connor and Alexandra Glazier},
journal={American Journal of Transplantation},
volume={21},
number={7},
pages={2325-2326},
year={2021}
}

@article{Jay,
title={{Measuring Transplant Center Performance: The Goals are Not Controversial but the Methods and Consequences Can Be}},
author={Colleen Jay and Jesse D. Schold},
journal={Current Transplantation Reports},
volume={4},
number={1},
pages={52-58},
year={2017}
}

@article{Efron,
title={{Large-Scale Simultaneous Hypothesis Testing: The Choice of a Null Hypothesis}},
author={Bradley Efron},
journal={Journal of the American Statistical Association},
volume={99},
number={465},
pages={96-104},
year={2004}
}

@article{efron2007size,
  title={{Size, Power and False Discovery Rates}},
  author={Efron, B.},
  journal={The Annals of Statistics},
  volume={35},
  number={4},
  pages={1351--1377},
  year={2007},
  publisher={Institute of Mathematical Statistics}
}

@book{efron2010large,
	title={Large-Scale Inference: Empirical Bayes Methods for Estimation, Testing, and Prediction},
	author={Efron, B.},
	year={2010},
	publisher={Cambridge, UK: Cambridge University Press}
}

@misc{SRTR_risk,
title={{SRTR Risk Adjustment Model Documentation: Waiting List Models}},
author={{Scientific Registry of Transplant Recipients}},
note={Available at \url{https://www.srtr.org/tools/waiting-list/}},
year={2021}
}

@misc{SRTR_psr,
title={{Technical Methods for the Program-Specific Reports}},
author={{Scientific Registry of Transplant Recipients}},
note={Available at \url{https://www.srtr.org/about-the-data/technical-methods-for-the-program-specific-reports/}},
year={2021}
}

@article{Proudlove,
author={Nathan C. Proudlove and Mhorag Goff and Kieran Walshe and Ruth Boaden},
title={{The Signal in the Noise: Robust Detection of Performance `Outliers' in Health Services}},
journal={Journal of the Operational Research Society},
volume={70},
issue={7},
pages={1102-1114},
year={2019}
}

@article{Wolfe,
author={Robert A. Wolfe and Valarie B. Ashby and Edgar L. Milford and Akinlolu O. Ojo and Robert E. Ettenger and Lawrence Y.C. Agodoa and Philip J. Held and Friedrich K. Port},
title={{Comparison of Mortality in All Patients on Dialysis, Patients on Dialysis Awaiting Transplantation, and Recipients of a First Cadaveric Transplant}},
journal={New England Journal of Medicine},
volume={341},
pages={1725-1730},
year={1999}
}

@article{Spiegelhalter_Plot,
author={David J. Spiegelhalter},
title={{Funnel Plots for Comparing Institutional Performance}},
journal={Statistics in Medicine},
volume={24},
issue={8},
pages={1185-1202},
year={2005}
}

@article{Shwartz,
author={Michael Shwartz and Joseph D. Restuccia and Amy K. Rosen},
title={{Composite Measures of Health Care Provider Performance: A Description of Approaches}},
journal={The Milbank Quarterly},
volume={93},
issue={4},
pages={788-825},
year={2015}
}

@article{Angrist,
author={Angrist, J.D. and Imbens, G.W. and Rubin, D.B.},
title={{Identification of Causal Effects Using Instrumental Variables}},
journal={Journal of the American Statistical Association},
volume={91},
issue={434},
pages={444-455},
year={1996}
}

@article{NM,
author={Nelder, J.A. and Mead, R.},
title={{A Simplex Method for Function Minimization}},
journal={The Computer Journal},
volume={7},
issue={4},
pages={308-313},
year={1965}
}

@book{Huber,
author={Huber, P.J.},
title={Robust Statistics},
publisher={New York: Wiley},
year={1981}
}

@book{Carlin,
author={Carlin, B.P. and Louis, T.A.},
year={2000}, 
title={Bayes and Empirical Bayes Methods for Data Analysis},
edition={2nd}, 
publisher={New York: Chapman \& Hall/CRC}
}

@article{Marrero,
author={Marrero, W.J. and Naik, A.S. and Friedewald, J.J. and Xu, Y. and Hutton, D.W. and Lavieri, M.S. and Parikh, N.D.},
title={{Predictors of Deceased Donor Kidney Discard in the United States}},
journal={Transplantation},
year={2017},
volume={101},
number={7},
pages={1690-1697} 
}

@Book{Venables,
    title = {Modern Applied Statistics with S},
    author = {W.N. Venables and B.D. Ripley},
    publisher = {Springer},
    edition = {Fourth},
    address = {New York},
    year = {2002}
}

\end{document}


\if0\blind
{
  \bigskip
  \bigskip
  \bigskip
  \begin{center}
    {\LARGE\bf Supplementary Material for \\ ``Composite Scores for Transplant Center Evaluation:
    A New Individualized Empirical Null Method"}
\end{center}
  \medskip
} \fi

\begin{center}
\end{center}
\smallskip
\begin{center}
\end{center}
\smallskip
\vfill

\newpage

\def\spacingset#1{\renewcommand{\baselinestretch}%
{#1}\small\normalsize} \spacingset{1}

\spacingset{1.0}
\section*{APPENDIX A}

We introduce the four measures of interest in a general form. Let $O_i$ and $E_i$ represent the observed and expected number of outcomes within the $i^{th}$ center, where the specific outcome for each measure is defined in Table A.1 and $E_i$ is calculated from the SRTR's risk-adjustment models \citep{SRTR_psr}. Each of the four measures is calculated as $O_i/E_i$. 

\begin{table}[h!]
    \centering
    \caption*{Table A.1. Description of outcome variable for each measure used in the composite score. TRR: Transplant Rate Ratio, SAR: Standardized Acceptance Ratio, PSMR: Patient Standardized Mortality Ratio, GMSR: Standardized Graft Failure Ratio.}
    \begin{tabular}{ccc}
    \hline
     \textbf{Measure}  & & \textbf{Outcome of Interest} \\
    \hline
    TRR & & Adult patient receives kidney transplant \\
    SAR & & Center accepts donated kidney for transplantation  \\
    PSMR & & Adult patient dies after kidney transplant \\
    GSMR & & Adult patient dies or has graft failure after kidney transplant \\
    \hline
    \end{tabular}
\end{table}

\section*{APPENDIX B}

\noindent Assume throughout that all expectations and variances are conditional on $\gamma^*_{ik}=0$ and the values of $\sum_{j=1}^{n_{ik}} b'_k(\theta^0_{ijk})$ and $\sum_{j=1}^{n_{ik}} b''_k(\theta^0_{ijk})$: \begin{align*}
    Var[Z_{FE,ik}]&=Var[E[Z_{FE,ik}|\alpha_{ik}]]+E[Var[Z_{FE,ik}|\alpha_{ik}]] \\
    &=Var\left[\frac{1}{\sqrt{a_k(\psi_k)\sum_{j=1}^{n_{ik}}b_k''(\theta^0_{ijk})}}\bigg \{E\left[\sum_{j=1}^{n_{ik}} Y_{ijk}\bigg |\alpha_{ik}\right]-\sum_{j=1}^{n_{ik}} b'_k(\theta^0_{ijk}) \bigg \} \right] \\
    & \quad+E\left[\frac{1}{a_k(\psi_k)\sum_{j=1}^{n_{ik}} b''_k(\theta^0_{ijk})}  Var\left[\sum_{j=1}^{n_{ik}} Y_{ijk}\bigg |\alpha_{ik}\right ]  \right] \\
    &=\frac{1}{a_k(\psi_k)\sum_{j=1}^{n_{ik}}b_k''(\theta^0_{ijk})}Var\left[\sum_{j=1}^{n_{ik}} \{b'_k(\theta^*_{ijk})-b'_k(\theta^0_{ijk})\}\right] \\
    & \quad + E\left[\frac{\sum_{j=1}^{n_{ik}} b''_k(\theta_{ijk}^*)}{\sum_{j=1}^{n_{ik}} b''_k(\theta_{ijk}^0)} \right].
\end{align*}

\noindent Taking Taylor series expansions of $b'_k(\theta^*_{ijk})$ and $b''_k(\theta^*_{ijk})$ around $\theta^0_{ijk}$, \begin{align*}
    b'_k(\theta^*_{ijk}) &\approx b'_k(\theta^0_{ijk})+b''_k(\theta^0_{ijk})\{\theta^*_{ijk}-\theta^0_{ijk}\} \\
    &=b'_k(\theta^0_{ijk})+b''_k(\theta^0_{ijk})\alpha_{ik},
\end{align*}

\noindent and $b''_k(\theta^*_{ijk}) \approx b''_k(\theta^0_{ijk})+b'''_k(\theta^0_{ijk})\alpha_{ik}$. Substituting these approximations into the original variance formula, \begin{align*}
    Var[Z_{FE,ik}]& \approx \bigg\{\frac{1}{a_k(\psi_k)} \sum_{j=1}^{n_{ik}} b''_k(\theta_{ijk}^0)\bigg\}Var[\alpha_{ik}]+1+\frac{\sum_{j=1}^{n_{ik}} b'''_k(\theta_{ijk}^0)}{\sum_{j=1}^{n_{ik}} b''_k(\theta_{ijk}^0)}E[\alpha_{ik}] \\
    &=1+\varphi_k \tilde{n}_{ik},
\end{align*}

\noindent where $\varphi_k=\sigma^2_{\alpha_k}/a_k(\psi_k)$ and $\tilde{n}_{ik}=\sum_{j=1}^{n_{ik}} b''_k(\theta_{ijk}^0)$.

\section*{APPENDIX C}

The MLE algorithm for estimating the $\varphi_k$ parameter in the empirical null variance is implemented as follows:

\begin{enumerate}
    \item Naive $Z_{FE,ik}$ values are computed using the fixed-effects score test, as described in Section 3.3.
    \item An intercept-only robust regression model is fit to the $Z_{FE,ik}$ values, using M-estimation with Tukey's biweight \citep{Huber,Venables}. 
    \item A sequence of $\pi_{0k}$ values is generated (e.g. from 0.8 to 1.0). For each  value of $\pi_{0k}$ in the sequence, the logarithm of the likelihood function from Section 3.5 is maximized with respect to $\varphi_k$, using the Nelder-Mead algorithm \citep{NM}. The estimated scale parameter from the robust regression model is used to compute the initial value of $\varphi_k$ in the Nelder-Mead algorithm. 
    \item The $\pi_{0k}$ value which yields the largest likelihood (evaluated at the corresponding $\widehat{\varphi}_k$ value) is selected as the MLE for $\pi_{0k}$. The corresponding $\widehat{\varphi}_k$ value is selected as the MLE for $\varphi_k$.
\end{enumerate} 

\section*{APPENDIX D}

\noindent For the Poisson model, $\tilde{n}_{ik}=\sum_{j=1}^{n_{ik}} \exp(\theta^0_{ijk})$. Assuming that $\alpha_{ik} \sim \textrm{N}(0, \sigma^2_{\alpha_k})$ and all expectations are conditional on the values of $\sum_{j=1}^{n_{ik}} b'_k(\theta^0_{ijk})$ and $\sum_{j=1}^{n_{ik}} b''_k(\theta^0_{ijk})$: \begin{align*}
    E[Z_{EN,ik}]&=E[E[Z_{EN,ik}|\alpha_{ik}]] \\
    &=E\left[\frac{\sum_{j=1}^{n_{ik}} \{ \exp(\theta^*_{ijk})-\exp(\theta^0_{ijk})\}}{\sqrt{\tilde{n}_{ik}(1+\sigma^2_{\alpha_k}\tilde{n}_{ik})}}\right] \\
    &=\sqrt{\frac{\tilde{n}_{ik}}{1+\sigma^2_{\alpha_k}\tilde{n}_{ik}}}\left[\exp(\gamma^*_{ik}) E[\exp(\alpha_{ik})]-1\right] \\
    &=\sqrt{\frac{\tilde{n}_{ik}}{1+\sigma^2_{\alpha_k}\tilde{n}_{ik}}}\left[\exp(\gamma^*_{ik}+\sigma^2_{\alpha_k}/2)-1\right]. \\
\end{align*} 

\noindent If two measures are combined such that higher Z-score values indicate better performances for the first measure and worse performances for the second measure, then the expected composite score is zero when \[\gamma^*_{i2}=log\left[\frac{\sqrt{\frac{\tilde{n}_{i2}}{1+\sigma^2_{\alpha_2}\tilde{n}_{i2}}}+\sqrt{\frac{\tilde{n}_{i1}}{1+\sigma^2_{\alpha_1}\tilde{n}_{i1}}}\left[\exp(\gamma^*_{i1}+\sigma^2_{\alpha_1}/2)-1\right]}{\sqrt{\frac{\tilde{n}_{i2}}{1+\sigma^2_{\alpha_2}\tilde{n}_{i2}}}\exp(\sigma^2_{\alpha_2}/2)}\right].\]

\renewcommand{\refname}{REFERENCES}

\bibliographystyle{JASA}
\bibliography{support}